\documentstyle[aps,prb,epsf]{revtex}

\def\batio3{BaTiO$_3$}

\begin{document}

\twocolumn[\hsize\textwidth\columnwidth\hsize\csname
@twocolumnfalse\endcsname

\title{Devonshire-Landau free energy of BaTiO$_3$ from
first principles}

\author{Jorge \'I\~niguez$^{1}$\cite{byline}, S. Ivantchev$^2$,
J. M. Perez-Mato$^2$, and Alberto Garc\'{\i}a$^1$}

\address{$^1$ Departamento de F\'{\i}sica Aplicada II, Universidad del
Pa\'{\i}s Vasco, Apdo 644, 48080 Bilbao, Spain\\ $^2$ Departamento de
F\'{\i}sica de la Materia Condensada, Universidad del Pa\'{\i}s Vasco,
Apdo 644, 48080 Bilbao, Spain}

\maketitle

\begin{abstract}
We have studied the Devonshire-Landau potential underlying the phase
transition sequence of BaTiO$_3$ using the first-principles effective
Hamiltonian of Zhong, Vanderbilt, and Rabe {[}{\it Phys. Rev. Lett.}
{\bf 73}, 1861 (1994){]}, which has been very successful in
reproducing the phase transitions and the dielectric and piezoelectric
properties of this compound. The configuration space (determined by
the polarization {\bf P} as order parameter) was explored with the
help of auxiliary electric fields. We show that the typically assumed
form of the potential, a sixth-order expansion in {\bf P} around the
paraelectric cubic phase, properly accounts for the behavior of the
system, but we find a non-trivial temperature dependence for all the
coefficients in the expansion, including the quadratic one, which is
shown to behave non-linearly. Our results also prove that the
sixth-order terms in the free-energy expansion (needed to account for
the first-order character of the transitions and the occurrence of an
orthorhombic phase) emerge from an interaction model that only
includes terms up to fourth order.
\end{abstract}

\vskip1pc
]

\narrowtext \marginparwidth 2.7in \marginparsep 0.5in

\section{INTRODUCTION}


From a phenomenological point of view, the behavior of a system in the
vicinity of a phase transition can be described in the framework of
Landau theory.~\cite{landau} In this scheme, one begins by identifying
the so-called order parameter, a (in general multidimensional)
variable ${\bf Q}$ that characterizes the symmetry change in the
transition, and then constructs the Landau free-energy function
$F({\bf Q};T)$, with the property that the equilibrium value of ${\bf
Q}$ as a function of temperature is that which minimizes
$F$. Formally, the Landau free energy is an incomplete thermodynamic
potential, which in principle could be calculated as:
\begin{equation}
F({\bf Q};T) = -k_BT\ln \sum_{j/{\bf Q}}{e^{-U_j/k_BT}} \; ,
\label{eq:incomplete}
\end{equation}
in analogy with the procedure to obtain the standard free energy in
terms of the partition function. Here the sum is only over those
states with the given value of ${\bf Q}$. In a Landau phenomenological
treatment, one just assumes that $F({\bf Q};T)$ exists, and that it
can be represented as a simple power series in ${\bf Q}$ in the
vicinity of the transition:
%
%
\begin{equation}
F({\bf Q};T) = F_0(T) + A(T)Q^2 + 
                        B\sum_{\beta}{f_{\beta}^{(4)}}({\bf Q})+ 
                        \dots \; ,
\label{eq:landau}
\end{equation}
where $f_{\beta}^{(j)}$ are invariants of order $j$ constructed from
${\bf Q}$. The quadratic coefficient $A$ is assumed to vary linearly
with temperature in the form $A(T)=\alpha(T-T_0)$, in such a way that
$T_0$ is the transition temperature (for a second-order transition) or
the lower metastability limit of the upper phase (for a first-order
one). These assumptions are known to break down in the critical region
close to the transition point of continuous phase transformations, but
they are valid when describing the approximate behavior of the system
in wider temperature intervals around the transition
temperature. Classical Landau theory ignores any temperature variation
of the higher-order coefficients in the expansion. Nevertheless, it
has been applied quite successfully to the analysis of many materials
in wide temperature ranges.~\cite{radescu} Nowadays Landau theory is
the most common approach to study the phenomenology of any
symmetry-breaking structural phase transition.

One of the early successes of this phenomenological approach to phase
transitions was the description by Devonshire~\cite{devonshire} of the
sequence of transitions in barium titanate (\batio3), a sequence that
spans a temperature interval of more than 200 degrees. This material
exhibits at high temperatures a paraelectric cubic perovskite
structure and, as temperature decreases, it undergoes three successive
first-order transitions to ferroelectric phases with tetragonal,
orthorhombic, and rhombohedral symmetries. In these phases the
polarization ${\bf P}$ points along one of the $\langle 1,0,0\rangle$,
$\langle 1,1,0\rangle$, and $\langle 1,1,1\rangle$ cubic directions
respectively.~\cite{jona} The polarization ${\bf P}$ can be identified
as the (three-dimensional) order parameter, and the Devonshire-Landau
potential (per unit volume) written as:
\begin{equation}
\label{eq-cowley}
\begin{array}{lll}
F & = & F_0+\frac{1}{2}aP^2 + u P ^4 + v (P_x^4+P_y^4+P_z^4)\\ 
      & & + h_1 P^6 + h_2 (P_x^6+P_y^6+P_z^6)\\ 
      & & + h_3 [P_x^4(P_y ^2+P_z^2)+P_y^4(P_z^2+P_x^2)\\
      & & + P_z^4(P_x^2+P_y^2)] \; ,
\end{array}
\end{equation}
where $F_0$ is the free energy of the reference cubic phase and $P^2$ stands
for $P_x^2+P_y^2+P_z^2$ (the notation is taken from the review paper by
Cowley~\cite{cowley}).  This is the complete expansion of $F$ up to
sixth order, and its relatively simple form is due to the high symmetry of the
system. Sixth-order terms in ${\bf P}$ in the expansion are needed to account
for the first-order character of the transitions (to model non-equivalent
coexisting free-energy minima).  Moreover, it can be seen that one needs
anisotropic sixth-order terms in $F$ to account for the orthorhombic phase of
\batio3.~\cite{vander-cohen} Devonshire showed that by assuming a linear
temperature dependence for $a$ and suitable {\sl constant} values for the rest
of the coefficients it is possible to qualitatively reproduce the transition
sequence as well as the dielectric properties of the system.~\cite{devon-more}


After Devonshire's work, the use of an expansion like that of
Eq.~\ref{eq-cowley} has become the standard way to model the
thermodynamic properties of ferroelectric perovskites, even though the
correctness of the assumption of constant high-order coefficients was
questioned early on. Drougard {\it et al.\/} and Huibregtse {\it et
al.\/}~\cite{exp} studied BaTiO$_3$ at temperatures just above the
cubic to tetragonal transition and in the orthorhombic phase
respectively; they assumed the existence of a Devonshire-Landau
potential but showed that there should be a significant temperature
dependence of the fourth-order terms.~\cite{gonzalo} It has since been
shown that there is a relatively large latitude to produce
Devonshire-like models that lead to qualitatively sensible predictions
for the transition sequence, divergence of the dielectric constants,
etc., even though these models may not be {\sl realistic} when confronted
with other experimental evidence.~\cite{other-forms} On the other
hand, it could be questioned whether such a relatively simple
free-energy expansion as that in Eq.~\ref{eq-cowley} exists at
all. For instance, Nakamura and Kinase have worked on a different
approach to the problem in which a non-polynomical from of the free
energy is used; they also discuss the connection of their work to the
Devonshire theory.~\cite{nakamura}

In the past decade, first-principles methods based on
density-functional theory have demonstrated to be accurate enough to
reproduce and predict structural phase transitions in perovskite
oxides.~\cite{vanderbilt} In particular, the phase transition sequence
and other properties of BaTiO$_3$,~\cite{zhong,garcia}
PbTiO$_3$,~\cite{waghmare-rabe} and KNbO$_3$~\cite{krakauer} have been
studied using the so called {\sl effective Hamiltonian} approach. An
ab-initio effective Hamiltonian $H_{\rm eff}$ is a mechanical model
that includes the relevant microscopic degrees of freedom of the
system and is constructed on the basis of first-principles
calculations. This model can then be analyzed by statistical methods
(typically, Monte Carlo or molecular dynamics) to explore the
finite-temperature behavior of the system. Monte-Carlo simulations
with an ab-initio effective Hamiltonian for \batio3\ have been
extremely successful, replicating approximately the experimental
transition sequence~\cite{zhong} and succeeding in reproducing the
main features of the dielectric and piezoelectric properties of the
real system.~\cite{garcia}

A natural question to ask then is to what extent the thermal behavior
resulting from these ab-initio effective Hamiltonians is compatible
with the Devonshire-Landau framework. In this paper we show that a
Devonshire-Landau expansion of the form of Eq.~\ref{eq-cowley} does
indeed emerge from detailed Monte Carlo simulations with the ab-initio
effective Hamiltonian for \batio3 of Ref.~\onlinecite{zhong}, with the
important qualification that the coefficients in the expansion have a
non-trivial temperature dependence. In particular, we show that the
sixth-order terms in $F({\bf P};T)$ are non-zero only in the
temperature range in which the transitions occur. Moreover, it becomes
clear that these sixth-order coefficients appear as
products of the statistical fluctuations of a Hamiltonian that only
includes terms up to fourth order in the polar degrees of freedom.

\section{METHOD AND TECHNICAL DETAILS}


Zhong, Vanderbilt, and Rabe~\cite{zhong} constructed their effective
Hamiltonian for BaTiO$_3$ retaining as relevant degrees of freedom of the
system a local polar distortion ${\bf u}_{i}$ in each cell, the homogeneous
strain $\eta$, and an inhomogeneous strain represented by a second set of
local vectors ${\bf v}_{i}$. The $H_{\rm eff}$ has the form:
\begin{equation}
\label{eq-zhong}
\begin{array}{lll}
H_{\rm eff} & = & \sum_{i,j;\alpha,\beta} J(i,j;\alpha,\beta)
u_{i\alpha} u_{j\beta} \\ & & + \sum_{i;\alpha,\beta}
\Gamma(\alpha,\beta) u_{i\alpha}^2 u_{i\beta}^2 \\ & & +
\sum_{i;\alpha,\beta;l} B(i;\alpha,\beta;l) u_{i\alpha} u_{i\beta}
\eta_{l} \\ & & + \sum_{l,h} B(l,h) \eta_{l} \eta_{h},
\end{array}
\end{equation}
where, for clarity, we have not written the terms associated to the
inhomogeneous strain. Here $i$ and $j$ range over the cells in the
system, $\alpha$ and $\beta$ are cartesian indexes, and $l$ and $h$
refer to tensor components in Voigt notation.  For the local polar
modes, the $H_{\rm eff}$ contains harmonic couplings
$J(i,j;\alpha,\beta)$ (on-site, $i=j$, and between modes in different
cells up to the third nearest neighbors) that reproduce the
instabilities of the cubic phase of \batio3,~\cite{uns-band} and
fourth-order on-site terms $\Gamma(\alpha,\beta)$ that define the
low-symmetry minima and stabilize the system. The effect of strain is
included through the standard elastic energy and through the coupling
coefficients $B(i;\alpha,\beta;l)$, which account for the spontaneous
strain in the low-symmetry phases. This model is probably the simplest
one that captures the essential physics of the system. As mentioned in
the introduction, $H_{\rm eff}$ is fourth-order in the polar
variables.

\begin{figure}
\epsfxsize=\hsize\epsfbox{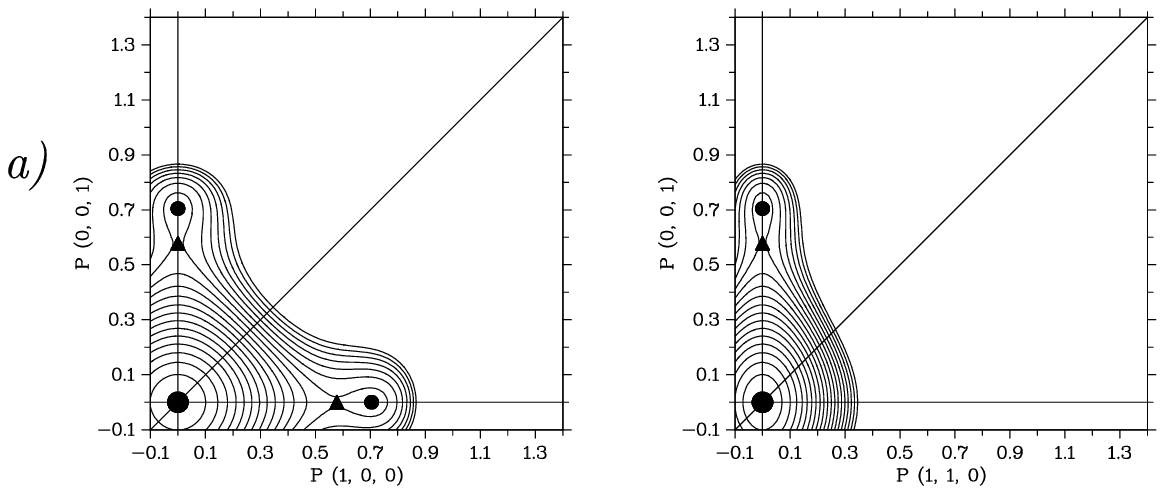}
\epsfxsize=\hsize\epsfbox{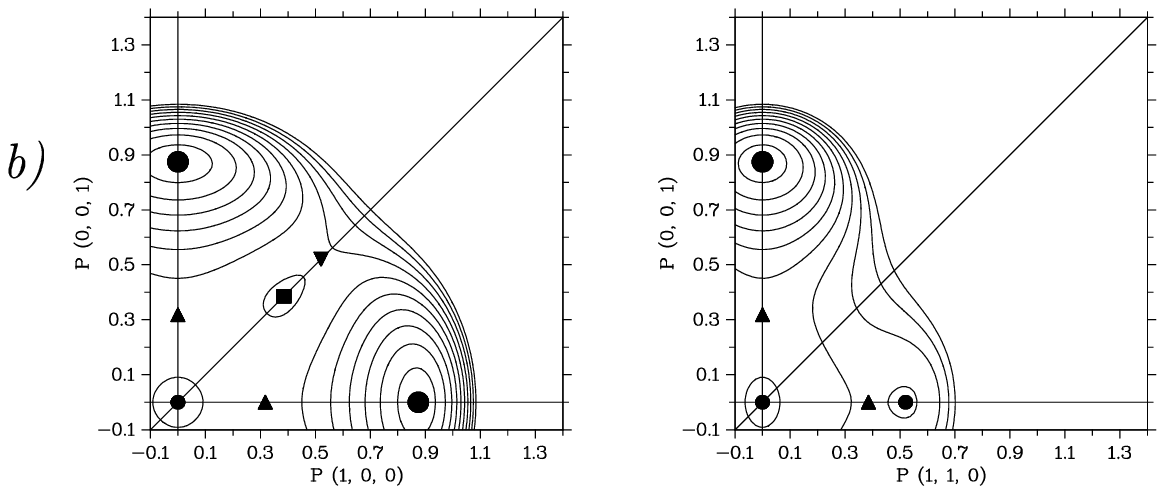}
\epsfxsize=\hsize\epsfbox{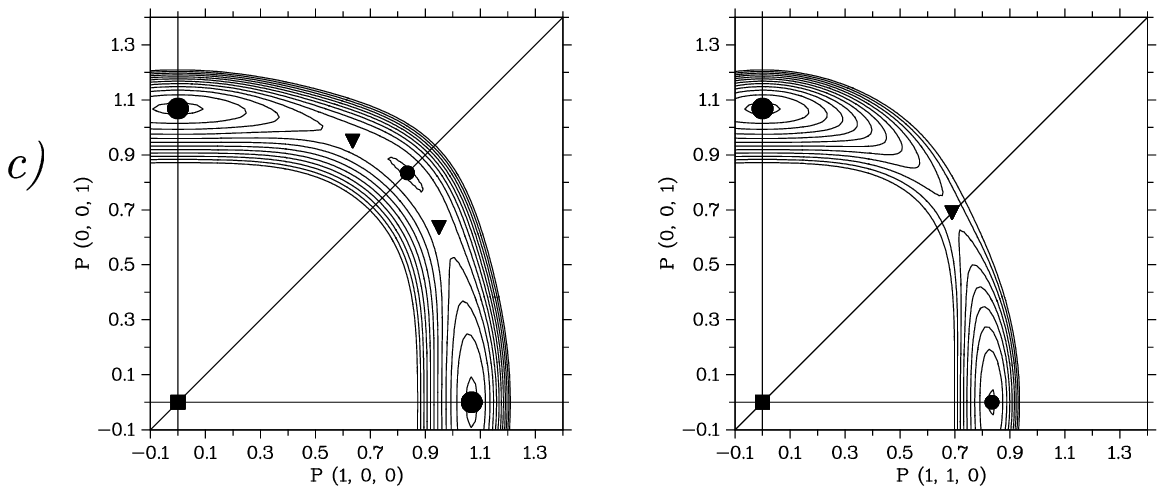}
\epsfxsize=\hsize\epsfbox{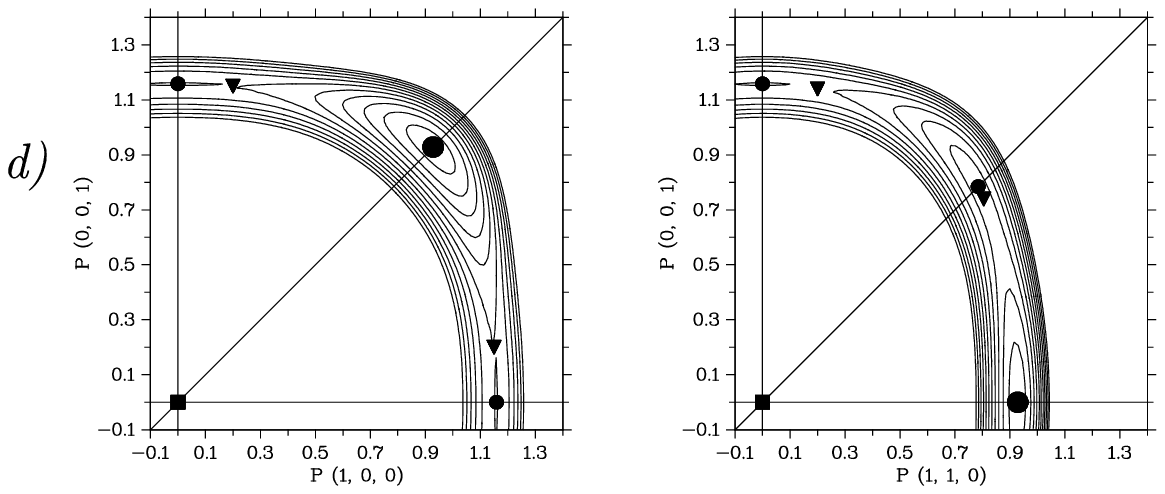}
\epsfxsize=\hsize\epsfbox{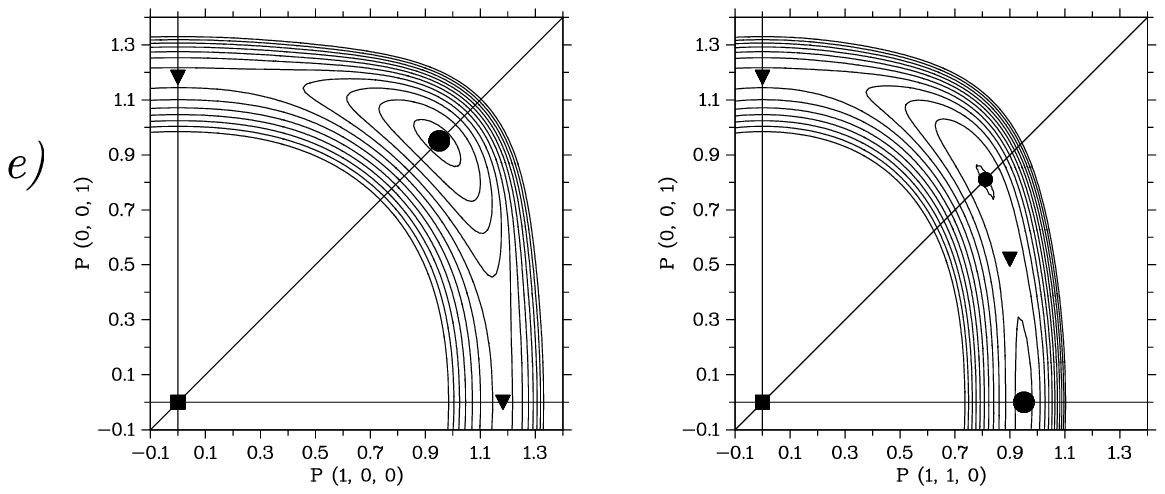}
\epsfxsize=\hsize\epsfbox{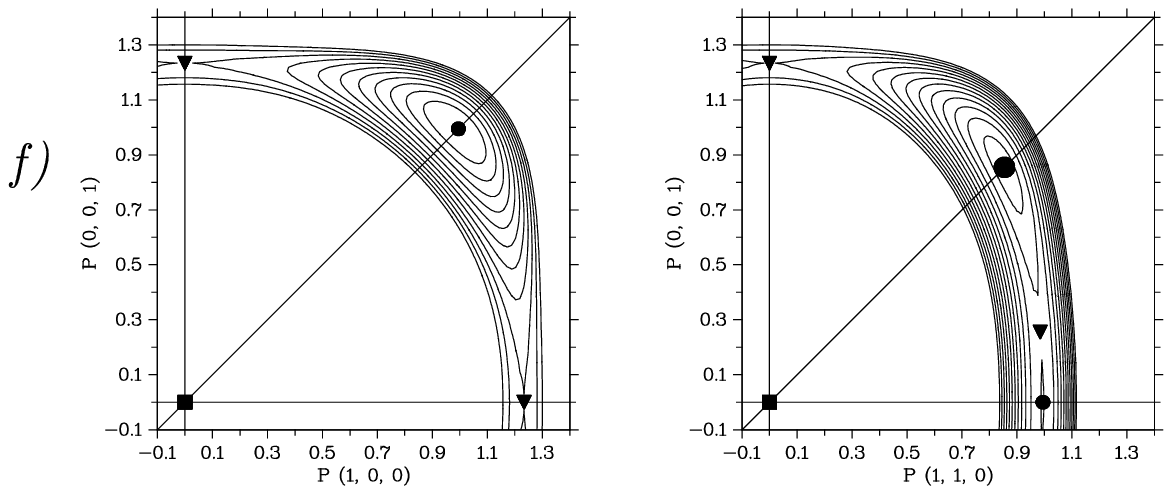}
\epsfxsize=\hsize\epsfbox{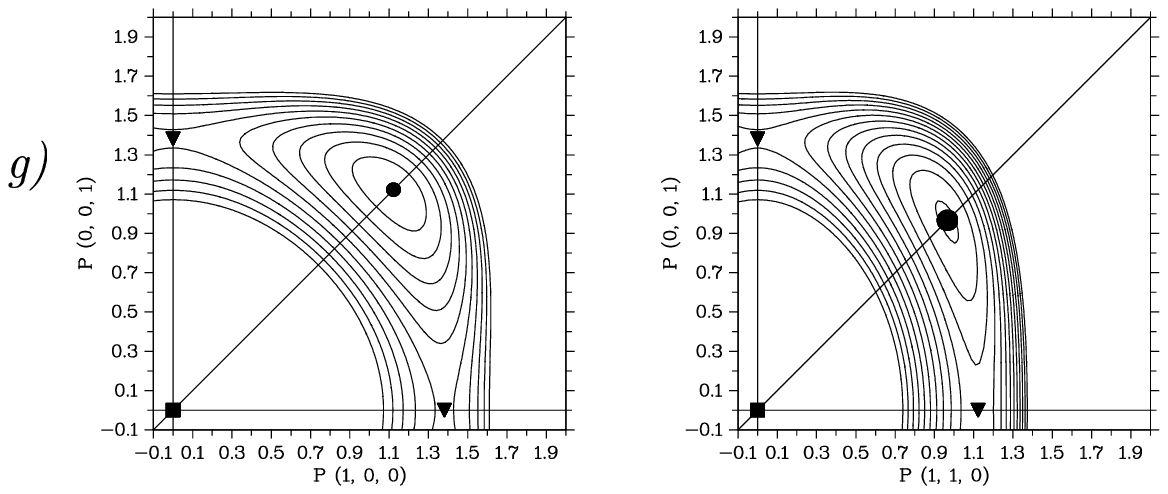}
\epsfxsize=\hsize\epsfbox{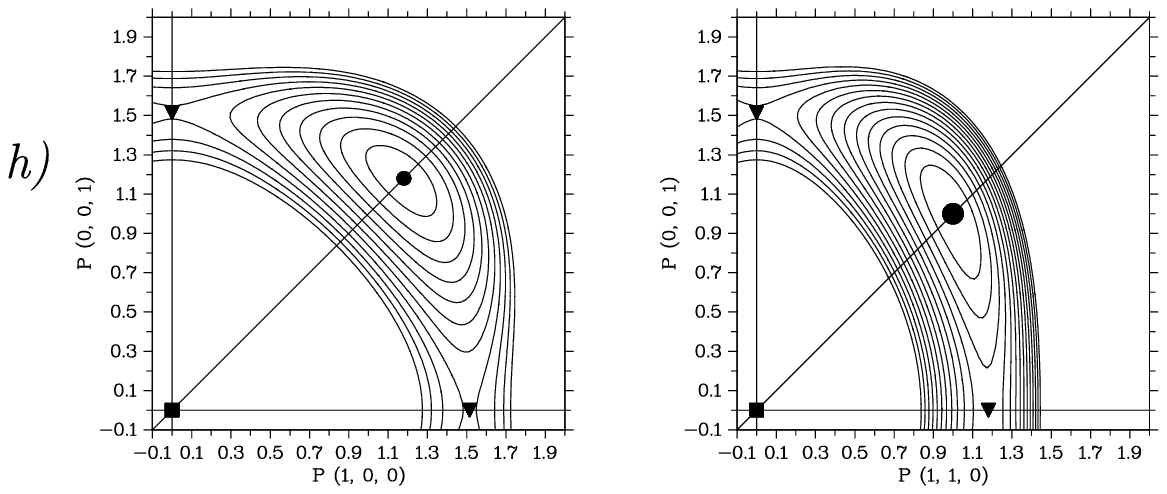}
\caption{Free-energy maps of BaTiO$_3$ at several temperatures. Panel
a): 300~K, b): 290~K, c): 250~K, d): 220~K, e): 210~K, f): 190~K, g):
100~K, and h): 0~K. For each temperature we show the (010) and
(1$\bar{1}$0) planes of the order-parameter configuration space, with
contour lines depicted only in the low free-energy regions. Symbols
characterize the relative stability of the critical points with
respect to displacements within a given plane: circles for free-energy
minima (a bigger circle corresponds to the stable-equilibrium state),
squares for free-energy maxima, and triangles for saddle points
(``up'' (resp. ``down'') triangles for points that are maxima
(resp. minima) along the radial direction). Note that critical points
with orthorhombic symmetry may need to be represented by two different
symbols in the $(010)$ and $(1\bar{1}0)$ planes (see panel b)).}
\label{fig-maps}
\end{figure}

A direct link between this $H_{\rm eff}$ and the phenomenological
potential of Eq.~\ref{eq-cowley} can be established in the very
low-temperature limit $T\rightarrow$ 0. Eq.~\ref{eq:incomplete} shows
that $F({\bf P};0)$ is just the energy of the lowest-lying state with
polarization ${\bf P}$. For \batio3 this state exhibits homogeneously
polarized cells with the global strain adjusted so as to minimize the
energy (the inhomogeneous strain is zero). It can be
shown~\cite{king-smith} that for a given ${\bf P}$, the homogenous
strain is proportional to the square of the polarization, so $F({\bf
P};0)$ is fourth-order in the polarization. This means that
$h_1=h_2=h_3=0$ in the low-temperature limit. Using the $H_{\rm eff}$
parameters, we obtain the rest of coefficients in $F$ at 0~K:
$a=-4/3$, $u=0.094$, and $v=0.051$. Here, units have been chosen so
that the first-octant rhombohedral ground state minimum is located at
${\bf P}=(1,1,1)$ and its energy per unit volume with respect to the
cubic phase is $-1$. (This election fixes the value of $a$ and the
relation $9u+3v=1$; i.e., $F({\bf P};0)$ is determined only by one
coefficient.) We have plotted this 0~K free-energy map in
Fig.~\ref{fig-maps}h), which shows the cubic phase as a local
free-energy maximum, tetragonal and orthorhombic saddle points (both
stable in the radial direction), and the rhombohedral global minima.

Now we tackle the calculation of $F$ at finite temperatures. A
straightforward approach to obtain information about the Landau
potential from the results of Monte Carlo~\cite{mc} simulations was
presented by Radescu {\sl et al.}~\cite{radescu} in the context of a
three dimensional $\Phi^4$ model.  From Eq.~\ref{eq:incomplete} it is
easy to show that there exists a relation between the probability
distribution of the order parameter ${\cal P}({\bf Q};T)$ and the
corresponding Landau potential $F({\bf Q};T)$:
\begin{equation}
\label{eq-radescu}
F({\bf Q};T) - F_{eq}(T) = -k_B T ln [{\cal P}({\bf Q};T)],
\end{equation}
where $F_{eq}(T)$ is the equilibrium free energy of the system at the
temperature considered. This procedure, which can be quite efficiently
applied to relatively simple cases such as the $\Phi^4$ model
(one-dimensional order parameter, a single second-order phase
transition), is not well suited for our problem. We have to study a
three-dimensional free-energy map with coexisting non-equivalent
minima. A typical Monte Carlo simulation gives information only about
one free-energy minimum (usually, the one corresponding to the
stable-equilibrium state of the system), with a poor sampling of the
order-parameter probability-distribution function in other regions. On
the other hand, more sophisticated sampling strategies would be very
demanding from the computational point of view.

This problem can be overcome, and $F({\bf P};T)$ computed in an
efficient manner, by modifying the effective Hamiltonian so as to
include the effect of an external electric field ${\bf
E}$:~\cite{garcia}
\begin{equation}
H'_{\rm eff} = H_{\rm eff} - {\bf E} \cdot Z^* \sum_i {\bf u}_i\; ,
\label{eq:hprime}
\end{equation}
where $Z^*$ is the Born effective charge associated to the local polar
distortions. As the quantity multiplying the electric field is just
the total dipole moment of the system, the new Landau potential is
simply $F'({\bf P};T,{\bf E}) = F({\bf P};T) - {\bf EP}$. An applied
electric field changes the location (and possibly the symmetry) of the
stable-equilibrium state. The new location can be determined by a MC
simulation of the modified effective Hamiltonian (${\bf
P}_{eq}=\langle {\bf P}\rangle$) and also computed
directly from $F'$. The idea can be mathematically expressed as:
\begin{equation}
\label{eq-def}
\left[\frac{\partial F'({\bf P};T,{\bf E})}{\partial {\bf P}}\right]_{{\bf
P}_{eq}(T,{\bf E})} = {\bf 0}\; .
\end{equation}
The left hand side of this equation depends linearly on the
coefficients of $F$ and on the electric field. Thus, for a given
temperature we can consider several electric fields, perform MC runs
to obtain the equilibrium values ${\bf P}_{eq}$,~\cite{tech1} and then
find the best solution of an overdetermined set of linear equations of
the form of Eq.~\ref{eq-def} to get the coefficients of the expansion
of $F$.

For each temperature, we first performed a zero-field calculation to
obtain a snapshot of the stable-equilibrium configuration. We then
used this configuration as the starting point for the runs with
electric field applied. In order to illustrate the kind of information
we were pursuing, consider $T$ such that the stable-equilibrium state
of the system is orthorhombic with ${\bf P}_{eq} = P_{eq} (1,1,0)$. In
this case, we would first use fields of the form $(E,E,0)$ to obtain
information about the response of the system within the orthorhombic
phase. Fields of the forms $(E,0,0)$, $(0,0,E)$, and $(E,E,E)$ would
probe the relative stability of minima with different
symmetries. Finally, general fields leading to $P_{x,eq} \ne P_{y,eq}
\ne P_{z,eq}\ne P_{x,eq}$, would explore in detail the intermediate
regions of the configuration space.  In our Monte Carlo simulations we
used a $14\times 14\times 14$ supercell (which corresponds to 13720
atoms) with periodic boundary conditions; we typically performed 15000
MC sweeps for the thermalization of the system and 35000 more to
calculate the averages of the polarization (these are very
well-converged calculation conditions, as can be checked in
Refs.~\onlinecite{zhong,garcia,waghmare-rabe}). For each temperature
we considered around one hundred different electric fields and
constructed a largely overdetermined system of equations for the
coefficients in $F$, which could then be reliably fitted.

\section{RESULTS AND DISCUSSION}

The data from our Monte Carlo simulations can indeed be represented by
a Landau free energy in the form of Eq.~\ref{eq-cowley} in a wide
temperature interval that includes all the transitions. (The
transition temperatures predicted by the effective Hamiltonian are
$T_{c1}=297$~K (cubic to tetragonal), $T_{c2}=230$~K (tetragonal to
orthorhombic) and $T_{c3}=200$~K (orthorhombic to rhombohedral)
respectively.~\cite{temps}) Our results show that the coefficients in
the expansion of $F$ have a significant and non-trivial temperature
dependence.

Before discussing in detail the behavior of the coefficients, it is
helpful to present the shape of the free energy they determine at a
few temperatures in the range of the phase
transitions. Fig.~\ref{fig-maps}a) (300~K) shows how tetragonal local
minima appear just above the first transition, while the absolute
minimum of $F$ is is still located at {\bf P}=0 in the cubic well.
Panels b) and c) correspond to 290~K and 250~K respectively, i.e.,
temperatures in the range of stability of the tetragonal phase.  We
see that the tetragonal wells are indeed the global free-energy
minima, and that the orthorhombic and rhombohedral wells nucleate in
the form of saddle points that are unstable with respect to a
tetragonal distortion (in the rhombohedral case, the saddle points are
unstable with respect to an orthorhombic distortion also).
Panels d) and e) refer to 220~K and 210~K respectively, i.e.,
temperatures in the orthorhombic range. The tetragonal wells become
metastable with respect to the orthorhombic ones, which are now the
free-energy global minima. At the same time, rhombohedral local minima
appear.
Finally, panels f) and g) correspond to 190~K and 100~K respectively,
both in the rhombohedral temperature range. Here we see that the
rhombohedral minima have finally become the deepest ones.
It is clear that the sequence converges to the free-energy map given
by the effective Hamiltonian itself (panel h)). For all the
transitions, the coexistence of different free-energy minima in the
figures evidences their first-order character.

Our method to explore the $F$ landscape has a limitation when probing
the vicinity of the cubic phase. After the point {\bf P}=0 becomes a
local free-energy maximum a few degrees below $T_{c1}$, it is no
longer possible to sample the region around it using auxiliary
electric fields: the cubic structure is already unstable and thus no
longer useful as a starting point for the simulations, and attempts to
steer a system in a tetragonal, orthorhombic, or rhombohedral well
towards {\bf P}=0 only succeed in landing it in the corresponding
symmetry-inverted domain. As the behavior of $F$ around {\bf P}=0 is
basically determined by the quadratic coefficient $a$ of the
Devonshire-Landau expansion, the value assigned to this coefficient by
a fit of the simulation data should be considered suspect. (This
problem pertains only to the region near {\bf P}=0. Information about
the relative stability of tetragonal, orthorhombic, and rhombohedral
phases is available at any temperature, since it is always possible to
apply fields that lead the system to each of these symmetries.)
\begin{figure}[b!]
\epsfxsize=\hsize\epsfbox{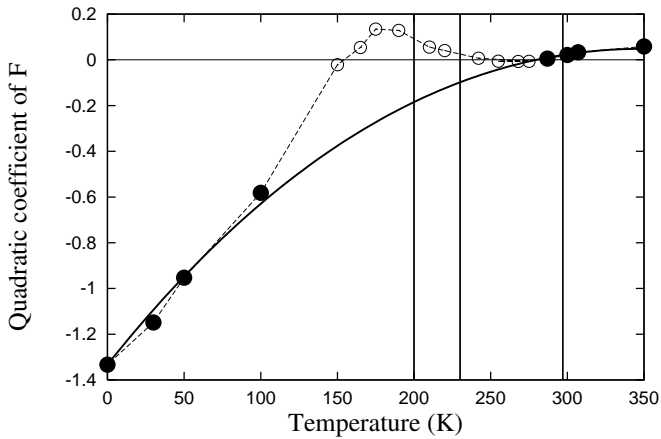}
\caption{Temperature behavior of the quadratic coefficient ($a$) of
the Devonshire-Landau potential: The dashed line connects the values
obtained from a direct fit to MC data, the solid line shows the
interpolated $a(T)$, and the solid circles are the points used to fit
the interpolating polynomial (see text). The transition temperatures
are marked with vertical lines.}
\label{fig-quadra}
\end{figure}
As shown in Fig.~\ref{fig-quadra}, at temperatures around and above
$T_{c1}$ the behavior of $a$ is fairly reasonable. It is positive at
high temperatures, becoming negative a few degrees below $T_{c1}$, as
corresponds to a first-order transition.  For very low temperatures
$a$ also behaves well, tending smoothly to its 0~K value. This is
because in this region the underlying Landau potential is so simple
(the thermal fluctuations are relatively small and the sixth-order
terms almost negligible) that we can use the information obtained
around the low-symmetry minima to reconstruct the complete free-energy
map. However, in the intermediate region (from 150~K to 250~K), $a$
turns positive again, which would mean that the cubic phase becomes
metastable in a wide temperature interval. This metastability is an
unphysical result (explicitly ruled out within our model by zero-field
MC simulations starting from a cubic phase, in which we found no trace
of it). We thus reconsidered the fitting of our data, imposing a more
physical temperature evolution of $a$: we assumed that in the
intermediate-temperature range $a(T)$ is given by a simple
interpolation (Fig.~\ref{fig-quadra}) between the high-temperature
region, where we can reliably sample it, and the low-temperature limit
determined directly by the convergence to the effective Hamiltonian (a
third-order polynomial suffices for our purposes). With this from for
$a(T)$ fixed, our MC data are still well fitted to the Landau
potential of Eq.~\ref{eq-cowley}. For instance, at 190~K the relative
error obtained when $a(T)$ is freely fitted is $2.2\%$, and when
$a(T)$ is fixed by the interpolation the error is still very small:
$3.6\%$. This clearly shows that, in the intermediate-temperature
region, our Monte Carlo data contain little information about the
value of $a$.~\cite{umbrella} Higher-order terms were tried but found
to play no role in the fit, so they can be excluded from the
potential. Our final results are shown in Fig.~\ref{fig-coeffs}.
\begin{figure}[b!]
\epsfxsize=\hsize\epsfbox{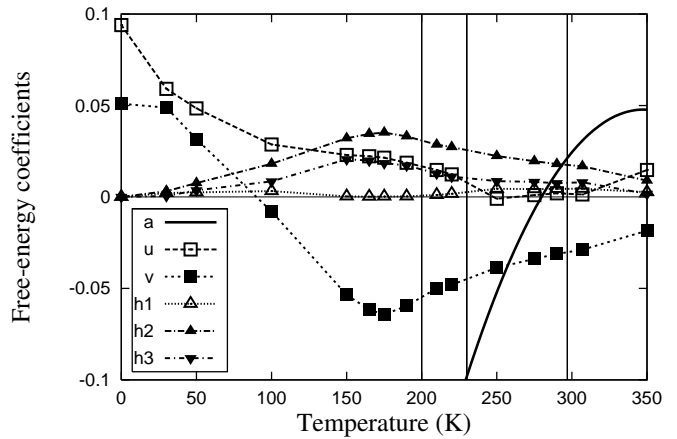}
\caption{Free-energy coefficients fitted to MC data assuming $a(T)$ is
given by the solid line in Fig.~\protect\ref{fig-quadra}. The
transition temperatures are marked with vertical lines.}
\label{fig-coeffs}
\end{figure}
At very low temperatures $a$, $u$, and $v$ tend to their 0~K values
and the sixth-order terms go to zero. On the other hand, at
temperatures well above the first transition ($T>350$~K) we find the
MC data are properly fitted to a fourth-order Landau potential, so the
sixth-order terms can again be excluded from the model. The sixth-order
terms then occur only in the temperature range in which the
transitions take place (as plainly shown in Fig.~\ref{fig-coeffs}). It
is also remarkable that the fourth-order terms exhibit a very strong
temperature dependence. Particularly, the anisotropic term $v$ changes
sign at around 90~K and takes large negative values all through the
intermediate-temperature range. So, apart from the influence it has,
together with $h_2$ and $h_3$, in determining the transition sequence,
we find that a negative $v$ is responsible for the first-order character of
the cubic to tetragonal transition.

\begin{figure}[b!]
\epsfxsize=\hsize\epsfbox{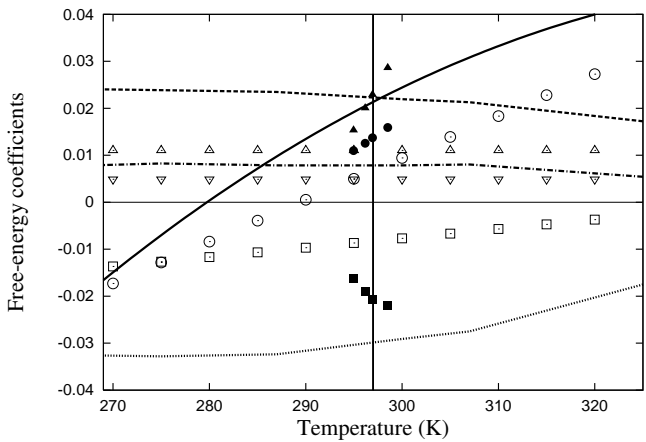}
\caption{Comparison of our free-energy coefficients (lines) with the
experimental values of Refs.~\protect\onlinecite{exp} (open symbols)
and~\protect\onlinecite{gonzalo} (solid symbols): solid line and
circles for $a$, dotted line and squares for $u+v$, dashed line and
``up'' triangles for $h_1+h_2$, and dash-dotted line and ``down''
triangles for $h_3$.  The vertical line marks the cubic-tetragonal
transition temperature.}
\label{fig-thexp}
\end{figure}

In Fig.~\ref{fig-thexp} we have plotted our coefficients near the
cubic to tetragonal transition temperature together with the available
experimental results.~\cite{exp,gonzalo} (We have changed the
temperature scale to make the experimental $T_{c1}$ coincide with the
theoretical one.) The agreement is only qualitative but as good as
could be expected given the simplifying assumptions involved in the
experimental work of Ref.~\onlinecite{exp} (the coefficients are
restricted to be constant or to depend linearly on temperature) and
the fact that the differences among the two experiments are comparable
to those between experiment and theory.

Our work strongly suggests that the phase transitions of BaTiO$_3$ can
be described in terms of a single Landau potential $F$ with the form
of Eq.~\ref{eq-cowley}. Despite what is assumed in most of the
previous work on this problem, the quadratic parameter $a$ is found to
exhibit a strongly non-linear temperature dependence. This is clear
because we can reliably calculate $a$ at high (over $T_{c1}$) and very
low temperatures, and the two regions cannot be joined linearly.  We
show that all the high-order terms in $F$ present a very significant
evolution with temperature. This conclusion is essentially opposed to
the temperature-independent behavior that is still assumed by some
authors (see, for example, Ref.~\onlinecite{other-forms}). It is also
very remarkable that the two features of BaTiO$_3$ that require the
inclusion of sixth-order terms in the Landau potential, i.e., the
first-order character of the transitions and the occurrence of an
orthorhombic phase, have been reproduced using a fourth-order
effective Hamiltonian.~\cite{PZT} This piece of information should be
taken into account when constructing mechanical models to study the
finite-temperature behavior of similar compounds. For instance, the
authors of Ref.~\onlinecite{nakamura} included sixth-order terms in
the underlying interaction model for \batio3\ with the explicit aim to
account for the first-order character of the transitions; our results
indicate that those terms should be unnecessary.

\section{SUMMARY}

We have studied the Devonshire-Landau potential underlying the phase
transition sequence of BaTiO$_3$ using the first-principles effective
Hamiltonian parametrized by Zhong, Vanderbilt, and Rabe. The
order-parameter configuration space was explored with the help of
auxiliary electric fields that change the location and relative
stability of the free-energy minima.  Our results show that the
typically assumed form of the potential, an expansion up to sixth
order in the polarization from the paraelectric cubic phase, properly
accounts for the behavior of the system. But, despite what is usually
presumed, we find a non-trivial temperature dependence for all the
coefficients in the expansion, including the quadratic term $a$, which
is shown to behave non-linearly. Our work also shows that the
sixth-order terms in polarization needed to explain basic features of
BaTiO$_3$ in a Devonshire-Landau approach (the first-order character
of the transitions and the occurrence of an orthorhombic phase) are
properly accounted for by an interaction model that only includes
terms up to fourth order.

\section{ACKNOWLEDGEMENTS}

This work has been supported by the Spanish CICyT grant
No. PB98-0244. J.~I. and S.~I. acknowledge financial support from the
Basque regional government and the UPV, respectively. We thank
Javier Junquera and Jon Saenz for their help.


\begin{thebibliography}{99}
\bibitem{mc} Although our emphasis is on Monte Carlo simulations,
the discussion applies also to molecular dynamics methods.
\bibitem{PZT} For Pb(Zr$_x$Ti$_{1-x}$)O$_3$, a similar fourth-order
effective Hamiltonian reproduces a monoclinic phase, and thus would
formally lead to eigth-order terms in the Devonshire-Landau
potential. See Ref.~\onlinecite{vander-cohen} and references therein.
\end{thebibliography}
\end{document}